\documentclass{iopart}

\usepackage{euscript,amssymb}
\usepackage{iopams}
\usepackage[dvips]{graphicx}
\catcode`\@=11
\renewcommand\footnoterule{%
  \kern-3\p@
  \hrule\@width.4\columnwidth
  \kern2.6\p@}
\renewcommand\@makefntext[1]{%
    \parindent 1em\noindent
    \hb@xt@1.8em{\hss$^{\@thefnmark}$)}\hspace{2pt}%
    \footnotesize\rmfamily#1}  
\def\@makefnmark{\hspace{.5pt}\hbox{$^{\@thefnmark}$%
\hspace{-1pt})}} 

\newcommand{\const}{\mbox{\rm const}\,}

\newcommand{\be}[1]{\begin{equation}\label{#1}}
\newcommand{\ee}{\end{equation}}
\newcommand{\ba}[1]{\begin{eqnarray}\label{#1}}
\newcommand{\ea}{\end{eqnarray}}
\newcommand{\rf}[1]{(\ref{#1})}

\def\RR{\mathbb{R}}

\def\ZZ{\mathbb{Z}}

\newcommand{\bj}{\mathbf{j}}

\newcommand{\bbr}{\mathbf{r}}

\newcommand{\bB}{\mathbf{B}}

\newcommand{\bH}{\mathbf{H}}

\newcommand{\cP}{\mathcal{P}}
\newcommand{\cT}{\mathcal{T}}

\def\p{\partial}
\def\a{\alpha}

\def\d{\delta}
\def\g{\gamma}

\def\sg{\sigma}
\def\e{\varepsilon}

\def\lb{\lambda}

\renewcommand\Im{\mbox{Im}}
\renewcommand\Re{\mbox{Re}}
\newcommand{\arctanh}{\mbox{\rm arctanh}\,}

\begin{document}

\title[A globally diagonalizable
$\alpha^2-$dynamo operator]{A globally diagonalizable
$\alpha^2-$dynamo operator, SUSY QM and the Dirac equation}

\author{Uwe G\"unther$^1$, Boris F Samsonov$^2$
and Frank Stefani$^1$}
\address{$^1$ Research Center Dresden-Rossendorf,  POB 510119,
D-01314 Dresden, Germany}
\address{$^2$ Physics Department, Tomsk State University, 36 Lenin Avenue,
634050 Tomsk, Russia}
\eads{\mailto{u.guenther@fz-rossendorf.de}, \mailto{samsonov@phys.tsu.ru} and \mailto
{f.stefani@fz-rossendorf.de}}

\begin{abstract}
A new class of semi-analytically solvable MHD $\a^2-$dynamos is
found based on a global diagonalization of the matrix part of the
dynamo differential operator. Close parallels to SUSY QM are used to
relate these models to the Dirac equation and to extract
non-numerical information about the dynamo spectrum.
\end{abstract}

The magnetic fields of planets, stars and galaxies are maintained
by dynamo effects in electrically conducting fluids or plasmas. A
crucial role in the qualitative understanding of the basic
dynamics of dynamos play various toy models which allow for a
semi-analytical study. In this respect, the spherically symmetric
$\alpha^2-$dynamo \cite{Moffatt,krause1} plays a role similar to
the harmonic oscillator in Quantum Mechanics (QM). But even this
model can be analytically described only in very few cases
--- like, e.g., for constant $\alpha-$profiles (The $\alpha-$profile
$\alpha(r)$ acts as an effective potential in the dynamo operator.)
or in the case of idealized boundary conditions \cite{GK-jpa2006}
(mimicking a perfect external conductor \cite{proctor1977} or the
limit of high angular mode numbers $l\gg 1$ \cite{large-L}).

Here, we are going to present another class of highly simplifying
$\alpha^2-$dynamo models based on a global diagonalizability of the
matrix part of the dynamo differential operator. For this purpose we
relax the very rigid boundary conditions (BCs) at the surface of the
dynamo maintaining fluid and replace it by a combination of a
strongly localized $\a$ embedded in a conducting surrounding
\cite{meinel} and Dirichlet BCs at infinity. This allows us to use
not only the Krein space symmetry properties of the operator
\cite{GK-jpa2006,GS-jmp1,GSZ-jmp2} (closely related to
$\cP\cT-$symmetric QM \cite{BB}) but also to uncover deep relations
to super-symmetric (SUSY) QM (for a recent review see
\cite{Mielnik}) and to map the dynamo eigenvalue problem into a set
of two $2\times 2$ stationary Dirac equations. Similar to
\cite{meinel}, we will find a close relation between overcritical
dynamo states and bound states in an associated QM model. For a
certain parameter value the diagonalization technique breaks down
and we develop a perturbative approach to describe the system
behavior in the vicinity of the corresponding operator Jordan
structure. The spectral reality properties of overcritical dynamo
states are discussed. Finally, we interpret our findings in terms of
a special link between the radial components of the electromotive
force and the induced currents.

Subject of our analysis is the eigenvalue problem of a spherically
symmetric $\alpha^2-$dynamo \cite{krause1} in its simplified and
unitarily re-scaled form \cite{GS-jmp1,GSZ-jmp2}
\be{i7}
\left(\partial_r  K\partial_r+ M-\lambda I\right)\Phi(r)=0\,, \quad
\Phi(0)=0\,,\quad \Phi(\infty)=0\,.
\ee
The matrix structure is encoded in
\begin{eqnarray}
 K&=&I-\alpha\sigma_-\,,\label{i5} \\
 M&=&- K\frac{l(l+1)}{r^2}+\alpha\sigma_+\,,\label{i6}
\end{eqnarray}
where $I$ is the $2\times2$ unit matrix and $\sigma_{\pm}$ denote
the nilpotent matrices $\sigma_\pm=\frac 12(\sg_x\pm i\sg_y)$ with
$\sg_{x,y,z}$ being the usual Pauli matrices. The system \rf{i7}
describes the coupled poloidal and toroidal components of the
$l-$modes of the magnetic field in a mean-field $\alpha^2-$dynamo
model with helical turbulence function ($\alpha-$profile) $\alpha
(r)$. We assume this $\alpha-$profile real-valued, bounded,
sufficiently smooth and exponentially decreasing for large $r\gg 1$.
The latter assumption allows us to consider the $\alpha-$profile
approximatively as localized and at the same time to relax the
otherwise rigidly imposed physical BCs at a given fluid/plasma
surface\footnote{The physics of MHD dynamos is discussed, e.g., in
\cite{Moffatt,krause1}.} (set, e.g., at $r=1$) replacing them by
Dirichlet BCs at $r\to\infty$. Such an approach will make underlying
structural links to quantum mechanical setups transparent and will
provide the Krein-space self-adjointness of the eigenvalue problem
\rf{i7} similar to that of models with idealized (Dirichlet BCs) at
fixed $r=1$ (see Refs. \cite{GK-jpa2006,GS-jmp1,GSZ-jmp2}).

Our first goal is in diagonalizing the matrix structure of the
$2\times 2$ matrix differential operator in \rf{i7}. For this
purpose we use a two-step procedure, which consists in a replacement
of the dependent variable by a Kummer-Liouville type transformation
\be{Af}
\Phi(r) = P^{-1}(r) \Xi(r)
\ee
to remove $\a\sigma_-$ \ from $K$ in the derivative term and an
afterwards performed coordinate-independent (global) similarity
transformation to diagonalize the remaining matrix potential. The
matrix $P$ can be found to be a square root of $K$,
\ba{k19}
&&K=P^2, \qquad P=\left(
                  \begin{array}{cc}
                    1 & 0 \\
                    -\frac \alpha 2 & 1 \\
                  \end{array}
                \right),
\ea
and yields the following equation for $\Xi(r)$:
 \be{Feq}
\Xi''(r)-\frac{l(l+1)}{r^2}\Xi(r) + V(r)\Xi(r)=0\,,
\ee
where
\be{V}
V(r)=\left(
                \begin{array}{cc}\frac 12 \alpha^2(r) -\lambda &\alpha(r)
                \\[.5em]
     \frac 12 \alpha''(r)+
     \frac 14\alpha^3(r)-\alpha(r)\lambda & \frac 12 \alpha^2(r) -\lambda
                \end{array}
      \right).
\ee
The equation system \rf{Feq} can be globally decoupled provided
the eigenvectors of the matrix \rf{V} are $r$-independent. It is
not difficult to see that this is possible if the function $\a(r)$
satisfies the equation
\be{Aeq}
\a''(r)+\frac 12\a^3(r)-a^2\a(r)=0
\ee
with $a$ an arbitrary real constant. The general solution to this
equation is expressed in terms of an elliptic integral which under
the additional requirement $|\alpha(r\to\infty)|\to 0$ reduces to
\be{Asol}
\alpha(r)=\frac{2a}{\cosh [a(r-r_0)]}
\ee
with $ar_0$ as an integration constant. For $\a-$profiles \rf{Asol}
the matrix part of the dynamo operator can be diagonalized
--- except for the special case $\lb=\frac 12a^2$ when it is similar
to a $2\times 2$ Jordan block. Under the diagonalization $V\mapsto
U^{-1}VU$ Eq. \rf{Feq} splits into the following decoupled pair of
differential equations (DEs) for the components $(F_+,F_-)$ of the
vector $U^{-1}\Xi=(F_+,F_-)^T$:
\be{FpmSyst}
\left[-\partial_r^2 +\frac{l(l+1)}{r^2} -\frac 12 \alpha^2\mp
\e\alpha\right]F_\pm =-\lambda F_\pm
\ee
with $\e=\left(\frac12 a^2-\lambda\right)^{1/2}$ or
$\e=-\left(\frac12 a^2-\lambda\right)^{1/2}$ and the diagonalizing
matrix $U$ given by
\be{U}
U=\left(%
\begin{array}{cc}
  1 & 1 \\
  \e & -\e \\
\end{array}%
\right)\,.
\ee
We notice that in equations \rf{Aeq} and \rf{FpmSyst} the parameter
$a$ is inessential and can be eliminated  by re-scaling $ \e=a
\tilde\e,\quad r=x/a,\quad \lambda=a^2\tilde \lambda,\quad
\alpha=a\tilde\alpha $. Further on, we will work in ``$a$ units"
what is equivalent to setting $a=1$ and identifying $r=x$,
$\e=\tilde \e$, $\lambda=\tilde \lambda$, $\alpha=\tilde\alpha$.
Apart from \rf{FpmSyst} we will also use reshaped versions of these
equations (obtained by substitution of $\lambda=\frac12-\e^2$) which
take the form of  quadratic pencils in the auxiliary spectral
parameter $\e$,
\be{1}\fl \qquad \ \ \
[-\p_x^2+\frac{l(l+1)}{x^2}-\frac 12\a^2+\frac
12\mp\e\a-\e^2]F_\pm=0\,, \quad \a=\frac{2}{\cosh(x-x_0)}\,,
\ee
supplemented by the Dirichlet BCs $F_\pm(x=0)=F_\pm(x=\infty)=0$.

In the special case $\e=0$, i.e. for $\lb=\lambda_J:=\frac 12$, the
diagonalization matrix $U$ (see \rf{U}) becomes singular and the
system \rf{Feq} assumes the upper triangular (Jordan-type) form
\be{EPeq}
\left(
  \begin{array}{cc}
    \partial_x^2-V_0 & -V_1 \\
    0 & \partial_x^2-V_0 \\
  \end{array}
\right)\left(
         \begin{array}{c}
           \Xi_1 \\
           \Xi_0 \\
         \end{array}
       \right)=0
\ee
with $\Xi_0$, $\Xi_1$ as components of the vector
$\Xi=(\Xi_1,\Xi_0)^T$ and the potential terms given by
$V_0=l(l+1)x^{-2}-\frac 12(\a^2-1)$, $V_1=-\a$.

We start our investigation of Eqs. \rf{FpmSyst}, \rf{1} by noticing
that for $l\neq 0$ they are DEs of non-Fuchsian type and therefore
their solutions cannot be expressed in terms of ordinary special
functions. A certain simplification occurs for the monopole case
$l=0$. Then \rf{FpmSyst}, \rf{1} constitute a particular type of
Heun's equations having three different finite regular singularities
(see e.g. \cite{Heun}) and solutions expressible in terms of Heun's
functions. A corresponding analysis will be presented elsewhere.

Here we are going to use the fact that Eqs. \rf{FpmSyst}, \rf{1} are
closely related to the exactly solvable stationary Schr\"odinger
equation
\be{sol1}
H_1\phi=E\phi\,,\qquad H_1=-\p_x^2-\frac12\alpha^2
\ee
with $H_1$ well known as superpartner\footnote{The potential
$-\alpha^2/2=-2/\cosh^2(x-x_0)$ is a $x_0-$shifted modified
P\"oschl-Teller \cite{grosche} and reduced Rosen-Morse potential
\cite{raab} and also known as "one-soliton potential well" in KdV
theory.} of the trivial Hamiltonian $H_0=-\p_x^2$
\ba{sol2}
&&LH_0=H_1L,\quad L=-\p_x+w',\quad w=\p_x\ln u,\quad
u=\cosh(x-x_0),\label{sol2-a}\\
&&L^\dagger L=H_0+1,\quad LL^\dagger =H_1+1,\quad (H_0-E_f)u=0,\quad
E_f=-1\label{sol2-b}
\ea
so that for $E>E_f=-1$ the solutions on the halfline
$x\in[0,\infty)$ are simply given as
\be{sol3}
\phi_\pm=L e^{\pm\kappa x}=\left[\mp
\kappa+\tanh(x-x_0)\right]e^{\pm\kappa x},\qquad \kappa=\sqrt{-E}\,.
\ee
Imposing the Dirichlet BC $\phi(x=0)=0$ and exponential decay on
these solutions makes $H_1$ an essentially selfadjoint operator with
a single discrete level $E(x_0)=-\tanh^2(x_0)\in (-1,0)$ for $x_0>0$
and with the halfline $E\ge 0$ as its continuous spectrum. The
additional dynamo-related constraint $\phi(x\to\infty)\to 0$ selects
then the bound state (BS) as relevant solution. With this
information at hand on the spectrum of the exactly solvable
Schr\"odinger equation \rf{sol1}, we are now well prepared to
present a qualitative discussion of the interrelated spectra of the
dynamo eigenvalue problems \rf{FpmSyst}.

A first piece of information can be extracted from \rf{FpmSyst} by
neglecting for a moment the potential terms $l(l+1)/x^2$ and
$\mp\e\a$\,. In this case \rf{FpmSyst} structurally coincides with
\rf{sol1} and we can identify $E=-\lambda$. This means that due to
the physical constraint $\phi(x\to\infty)\to 0$ and its implication
$E\in (-1,0)$ the model necessarily describes overcritical dynamo
regimes\footnote{A dynamo in its kinematic regime is called
overcritical when it has a positive growth rate $\Re(\lambda)>0$.
This is in contrast to so called undercritical regimes
$\Re(\lambda)<0$ with decaying (dissipating) magnetic field. (See
e.g. \cite{Moffatt,krause1}.)} $\lambda>0$.

\begin{figure}[th]
\begin{center}
\begin{minipage}{0.45\textwidth}
\includegraphics[angle=0, width=0.9\textwidth]{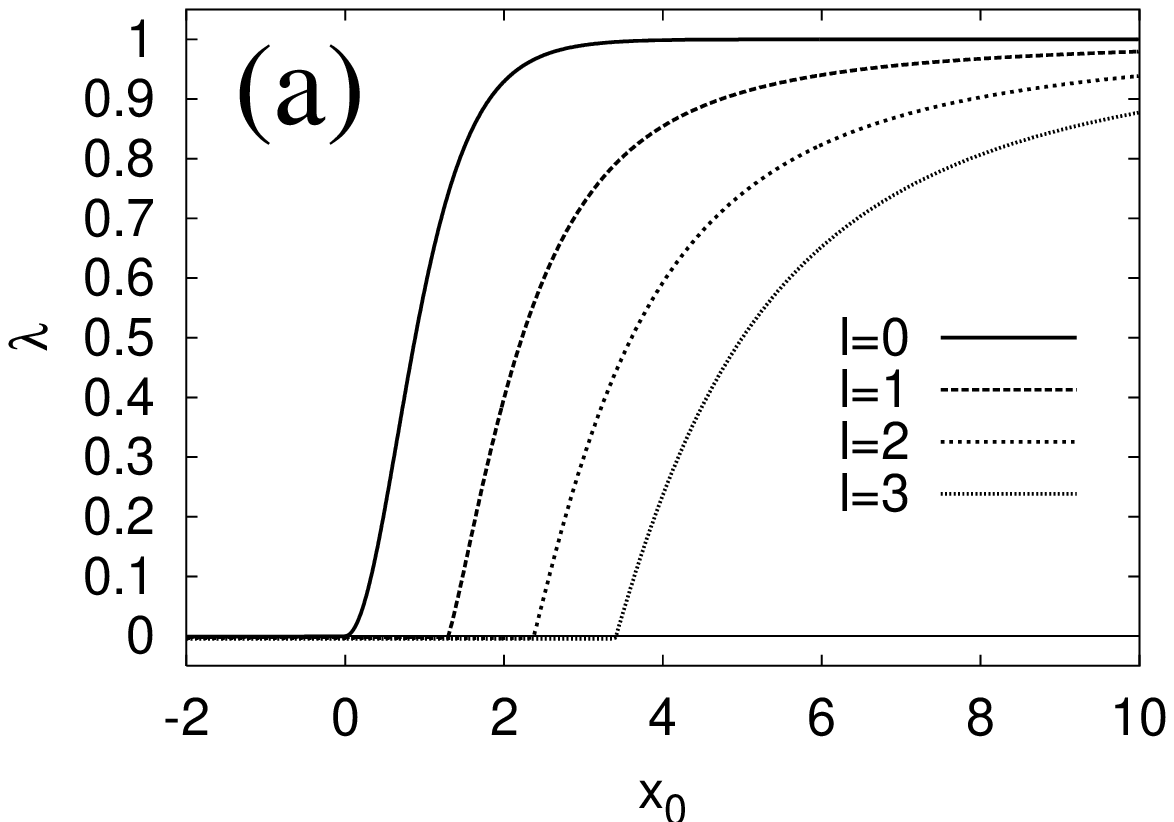}
\end{minipage}
\begin{minipage}{0.45\textwidth}
\includegraphics[angle=0, width=0.9\textwidth]{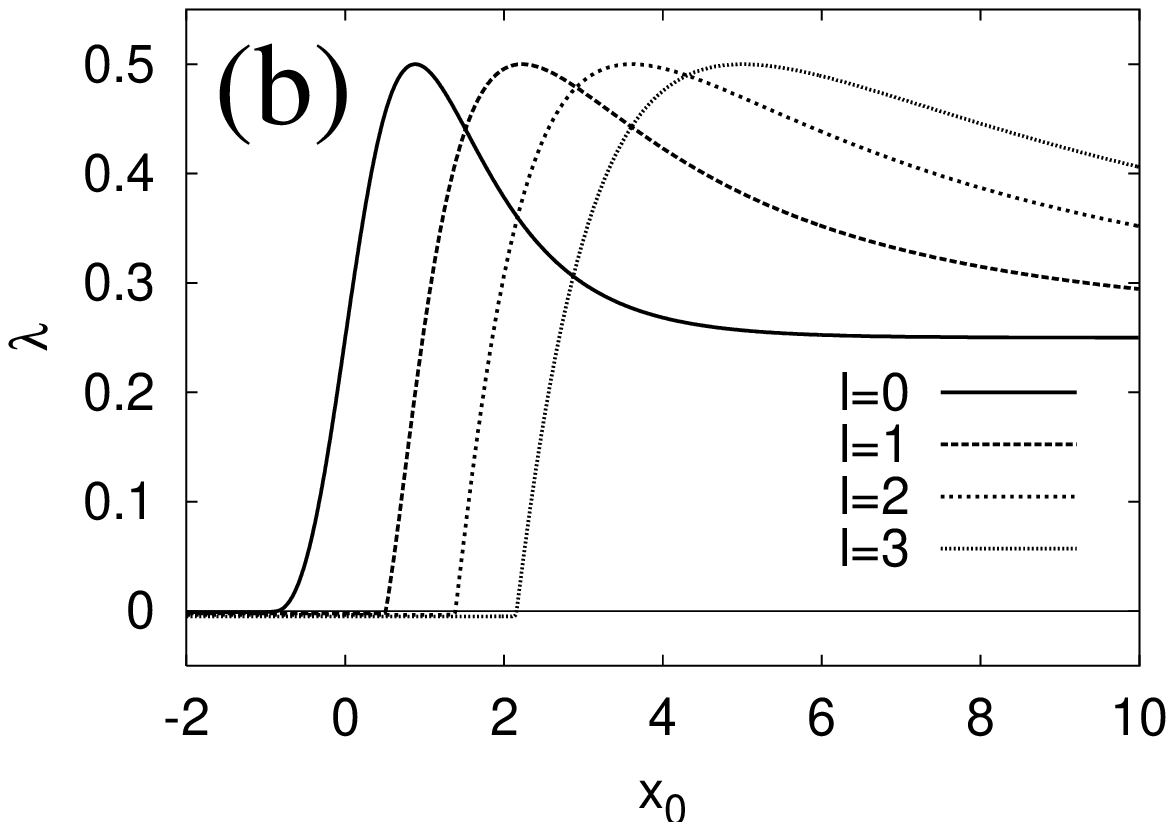}
\end{minipage}
\end{center}
\caption{\label{fig1}\small Spectra $\lambda(x_0)$ for the reduced
system $[-\p_x^2+l(l+1)/x^2-\alpha^2/2]F_\pm=-\lambda F_\pm$ (a) and
for the complete problem \rf{FpmSyst} (b) in case of angular mode
numbers $l=0,1,2,3$. For numerical reasons the Dirichlet BC has been
imposed at the large distance $x=100$.}
\end{figure}

In order to extract further information, we proceed with the
familiar QM model and extend $H_1$ by the centrifugal potential
$l(l+1)/x^2$. This potential acts as a repulsive barrier in the
vicinity of the origin $x=0$ and for small $x_0$ it overcompensates
the effect of the attractive one-soliton potential well $-
\alpha^2/2=-2/\cosh^2(x-x_0)$ with center at $x=x_0$. As result, no
BS can exist for small $x_0$. When $x_0$ is increased beyond a
certain $l-$dependent critical value the effect of the repulsive
barrier will become sufficiently weak and the BS level will
re-appear from the lower boundary $E=0$ of the continuous spectrum
and move down toward the lower boundary $E\to E_f=-1$ of the BS
band. The situation is illustrated in Fig. \ref{fig1}a \ showing the
sign-inverted picture for $\lambda=-E$.

It remains to clarify the role of the energy dependent part
$\left.\mp\e\a=\mp(\frac12-\lambda)^{1/2}\alpha\right.$ of the
potential. Due to the specific square-root coupling this term might,
in general, become complex-valued. Hence, the main question to
answer is whether the spectrum will remain purely real (as the
numerical results in Fig. \ref{fig1}b \ indicate) or whether and
under which conditions it might become complex.

A partial clarification can be achieved by transforming the pencil
equations \rf{1}  into equivalent Dirac equations. For this purpose
we use a more general and slightly reshaped version of the SUSY
factorization technique \rf{sol2-b} (cf. \cite{Dirac-Miura}),
factorizing instead of $H_1$ the Hamiltonian
\ba{sol11}
&&H_{2,l}=-\p_x^2+\frac{l(l+1)}{x^2}-\frac12\alpha^2+\frac12=L^\dagger L\,,\label{sol11-a}\\
&&L=-\p_x+w,\quad L^\dagger=\p_x+w,\quad w=u'/u\,,\label{sol11-b}
\ea
which has a continuous spectrum for $E\ge 1/2$ and a BS at some
$E<1/2$. The factorization \rf{sol11-a} allows us to rewrite the
pencils \rf{1} as
\be{1a}
[L^+L\mp\e\a-\e^2]F_\pm=0\,.
\ee
Denoting now $F_\pm=:\psi_1$ and assuming for $\e\ne0$
\be{De1}
L\psi_1=:\e\psi_2
\ee
equation (\ref{1a}) can be expressed as
\be{De2}
L^+\psi_2-(\pm\a+\e)\psi_1=0\,.
\ee
From the explicit form (\ref{sol11-b}) of the operators $L$ and
$L^+$ we see that equations (\ref{De1}) and (\ref{De2}) are nothing
but one-dimensional Dirac systems in their general representation
(see e.g. \cite{kost})
\be{DS}
H\Psi=\e\Psi\,,\quad
H=\g\p_x+V
\ee
with\footnote{With the help of a gauge transformation $\Psi=A\Theta$
(see e.g. \cite{kost}) which does not affect the zero boundary
conditions, the Dirac equations \rf{DS} can be easily transformed
into their canonical forms corresponding to a combination of scalar
and pseudoscalar fields and describing the movement of massless
particles.}
\be{DH}
\Psi=\left(
\begin{array}{c}
\psi_1\\\psi_2
\end{array}
\right),\quad
\g=\left(
\begin{array}{cc}
0 & 1\\-1 & 0
\end{array}
\right),\quad
V=\left(
\begin{array}{cc}
\mp\a & w\\w & 0
\end{array}
\right).
\ee
For the factorization \rf{sol11-a} to work, the function $u(x)$
should be a solution of the equation $H_{2,l}u(x)=0$, i.e. an
eigenfunction at the factorization energy $E_f=0$. At the same time
it should be nodeless on the positive semiaxis providing in this way
a superpotential $w(x)$ which is regular. According to an
implication of the ``oscillation" theorem (see, e.g.,
\cite{Berezin}) a real eigenfunction of the operator \rf{sol11-a}
can be nodeless only if its eigenvalue is located below the ground
state level if the latter exists or below the lower bound of the
continuous spectrum otherwise. Applied to our configuration with
$E_f=0$ this means that $u$ is nodeless as long as the BS (ground
state) energy $E(x_0)>E_f$. Comparison of \rf{sol11-a} with \rf{1}
shows that $E_f=0$ coincides with the energy of the Jordan
configuration \rf{EPeq} at $\e=0$ so that for $x_0<x_J$ the solution
$u$ is nodeless and therefore $w$ continuous. Together with the
Dirichlet BC $\psi_1(x=0)=0$ and the easily verified boundedness of
$|\psi_2(x=0)|=|\e^{-1}L\psi_1|_{x=0}\le C<\infty$ for $\e\neq 0$
this makes an adapted version of Theorem III.7.1 from \cite{kost}
applicable: If the coefficients of a Dirac system are continuous
functions in any finite interval of the positive semi-axis then the
Dirac operator $H$ defined by the differential system (\ref{DS}) and
the boundary condition $\psi_1(0)\cos\d+\psi_2(0)\sin\d=0$ (with
$\d$ an arbitrary real number) is selfadjoint provided its domain of
definition is $\psi_1,\psi_2\in{\cal L}^2(0,\infty)$ with $
L\psi_1\in{\cal L}^2(0,\infty)\,,\quad L^+\psi_2-\a\psi_1\in{\cal
L}^2(0,\infty)\,. $ This  means that the Dirac system \rf{DS}
supplemented with the Dirichlet BC at the origin, and with it the
dynamo problem, has a purely real spectrum for $x_0< x_J$.

In case of $x_0>x_J$ any real-valued solution to equation
$H_{2,l}u=0$ has a node on the positive semi-axis so that the
superpotential $w(x)$ has a pole at some $x>0$ and the theorem is no
longer applicable. A circumvention of this problem might consist in
a complex linear combination $u(x)=u^{(1)}(x)+icu^{(2)}(x)$,
$c\in\RR$ of two real-valued linearly independent solutions
$u^{(1)}$ and $u^{(2)}$ of equation $H_{2,l}u=0$. Such a function
$u$ has no nodes on the positive semi-axis, but the superpotential
$w$ becomes a complex-valued function. The spectrum of such a Dirac
operator needs a special analysis which will be presented elsewhere.

\begin{figure}[th]
\renewcommand{\thefigure}%
{\arabic{figure}}
\begin{center}
\includegraphics[angle=0, width=0.4\textwidth]{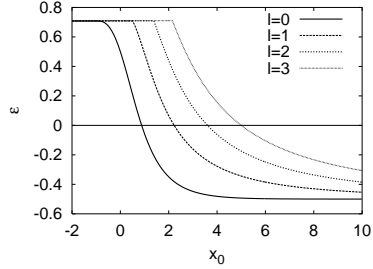}
\caption{\label{fig3}\small\noindent Spectrum in terms of $\e(x_0)$
for angular mode numbers $l=0,1,2,3$.}
\end{center}
\end{figure}

Here we proceed with a general qualitative analysis.  We start from
the observation that for small real $\lambda\le 1/2$ the square root
$\e$ is real. Introducing an auxiliary parameter $b\in\RR$ and
replacing the potential term $\left.\mp\e\a\right.$ in \rf{FpmSyst}
by $b\a$ the quadratic pencils \rf{1} in $\e$ reduce to an auxiliary
linear eigenvalue problem in a $b-$dependent $\lambda(x_0,b)$ with
additional constraints $b=\mp\e=\mp\left(1/2-\lambda\right)^{1/2}$.
The existence problem of the pencil solutions can be easily studied
graphically in the $(b,\lambda)-$plane. A solution exists if the
plot of the numerically obtained $\lambda(x_0,b)$ for given $x_0$
has an intersection point with the graphics of one of the
constraints $b=b(\lambda)$. The corresponding analysis shows that
for the branch $\e\ge 0$ one has a BS with $(F_+\not\equiv
0,F_-\equiv 0)$ for $x_0<x_J$ and with $(F_+\equiv 0,F_-\not\equiv
0)$ for $x_0>x_J$. Due to the invariance of \rf{FpmSyst} and \rf{1}
under the sign (branch) change $(\e,F_\pm)\mapsto (-\e,F_\mp)$ this
can be re-interpreted as a single solution $(F_+\not\equiv
0,F_-\equiv 0)$ with $\e(x_0<x_J)>0$ and $\e(x_0>x_J)<0$. The latter
interpretation is confirmed by a direct numerical analysis (see Fig.
\ref{fig1}b) of Eqs. \rf{FpmSyst} (cross-checked by numerics on the
original matrix-operator problem \rf{i7}) and by the graphics of
$\e(x_0)$ in Fig. \ref{fig3}.
Although a rigorous and complete analytical proof of the  result for
all $x_0\in \RR,\ l\ge 0$ is still missing, we now present a
perturbative treatment of an $l=0$ model in the vicinity of the
non-diagonalizable operator-configuration \rf{EPeq} at $\e=0$ which
explains the numerical results at least locally.

For $l=0$ the Jordan chain equations \rf{EPeq} have the form
\ba{sol4}
\left[-\p_x^2-\a^2/2+1/2\right]\Xi_0=0,\quad
\left[-\p_x^2-\a^2/2+1/2\right]\Xi_1=\a \Xi_0
\ea
and due to their relation to the eigenvalue problem \rf{sol1},
$(H_1-E)\phi=0$, $E=-1/2$ they are exactly solvable. From \rf{sol3}
we immediately read off that for BSs
\be{sol5}
\Xi_0=C_0\phi_-\,,\qquad \kappa=2^{-1/2},\qquad C_0=\const.
\ee
At the same time we find the value of the parameter $x_0=x_J$ for
which the Jordan structure occurs: $E(x_J)=-1/2=-\tanh^2(x_J)$
yields $x_J=\arctanh(2^{-1/2})$. The second (associated) BS
component $\Xi_1$ is easily obtained by standard techniques for
inhomogeneous ODEs (see, e.g., \cite{polyanin-DE})
\be{sol6}
\Xi_1=C_1\phi_-+\frac{C_0}W\left[\phi_+\int_0^x\a\phi_-^2dx'-\phi_-\int_0^x\a\phi_+\phi_-dx'\right],
\ee
with $W$ denoting the constant Wronskian $W\equiv
W(\phi_+,\phi_-)=-2^{-1/2}$. In the limit $x\to\infty$  the product
$\phi_+\int_0^x\a\phi_-^2dx'$ diverges so that the BS condition
$\Xi_1(\infty)=0$ implies $C_0=0$ and the full Jordan chain solution
reads simply $\Xi=(\Xi_1,\Xi_0)^T=(C_1\phi_-,0)^T$. Comparing the
chain equations \rf{sol4} at $\Xi_0=0$ with the pencil equations
\rf{1}, i.e. with
\be{sol7}
[-\p_x^2-\frac 12\a^2+\frac 12\mp\e\a-\e^2]F_\pm=0\,,
\ee
shows that for BSs Eqs. \rf{sol4} can be interpreted as effective
smooth limit of \rf{sol7} at $\e\to 0$. Hence, a BS perturbation
theory can be constructed simply on the decoupled scalar Eqs.
\rf{sol7} alone. As small perturbation parameter we choose the
$x_0-$distance $\delta=x_0-x_J$ from the Jordan configuration and
expand $\alpha=\alpha_J-\alpha'_J\delta+\ldots$,
$\e=e_1\delta+e_2\delta^2+\ldots$,
$F_\pm=\Xi_1+\chi_\pm\delta+\ldots$, where $\alpha_J=2/\cosh(x-x_J)$
and $e_1$, $e_2$, \ldots are coefficients to be defined from the
perturbation scheme. This yields the defining equation for the
first-order corrections $\chi_\pm$
\be{sol8}
\left[-\p_x^2-\a^2/2+1/2\right]\chi_\pm=-g_1\Xi_1,\qquad
g_1:=\a_J[\a_J'\mp e_1]
\ee
with solutions
\be{sol9}
\chi_\pm=C_\pm \phi_-+\frac{C_1}{W}\left[\phi_-\int_0^x
g_1\phi_+\phi_- dx'-\phi_+\int_0^x g_1\phi_-^2 dx'\right]\,.
\ee
The BS condition $\chi_\pm(x\to\infty)\to 0$ can only be fulfilled
if $\int_0^\infty g_1\phi_-^2 dx'=0$. Hence, it fixes the parameter
\be{sol10}
e_1=\pm\frac{\int_0^\infty\alpha_J\alpha_J'\phi_-^2
dx'}{\int_0^\infty\alpha_J\phi_-^2 dx'}=\mp\frac12
\ee
and with it $\e=\mp\delta/2$. Knowing that for fixed $x_0$ the
spectral parameter $\e$ is the same in the equations for $F_+$ and
$F_-$ we conclude that $\e=\mp\delta/2$ can be valid only for one of
the signs and that therefore it acts as a selection rule. Full
compatibility with the numerical results and Fig. \ref{fig3} is
established by choosing the $F_+-$related BS for $\e>0$ and
$x_0<x_J$, i.e. for $\delta<0$, so that $\e=-\delta/2$ holds for all
sufficiently small $\delta \in (-c_1,c_2)$ and provides a smooth
connection between the $\e>0$ and $\e<0$ branches. At the same time
it excludes a BS for the solution $F_-$\,.

Inspection of the recurrence algorithm for the higher order
corrections shows that at each order the highest-order coefficient
$e_k$ enters its defining equation only linearly so that no square
roots are involved which could produce complex-valued contributions.
Together with the reality of all other ingredients
($\phi_\pm,\alpha_J,\alpha_J',\ldots$) of these recurrence equations
we conclude that $\Im(e_k)=0,\ \forall k\in\ZZ^+$ and no
complex-valued BS-$\e$  can emerge from an BS within the convergence
region of the series $\e=\sum_{k=1}^\infty e_k \delta^k$. In this
way we found an argumentation complementary to the Dirac equation
based technique for $x_0<x_J$. Another argument explaining the
reality of the BS eigenvalue follows from the fact that the dynamo
operator \rf{i7} with Dirichlet BCs is necessarily self-adjoint in a
Krein space \cite{GK-jpa2006,GS-jmp1,GSZ-jmp2}, and, hence, a
spectral real-to-complex transition requires two spectral branches
of different Krein-space type to coalesce at some point in parameter
space. Once locally only a single BS exists there is also no chance
for a BS-related spectral phase transition to complex eigenvalues
within the convergence region of $\e=\sum_{k=1}^\infty e_k
\delta^k$. The question of whether complex-conjugate BS eigenvalue
pairs might split off from the continuum remains still open.

Finally we interpret the obtained solution behavior $F_+\not\equiv
0$, $F_-\equiv 0$ in terms of a special link between the magnetic
field components of the dynamo. According to \cite{Moffatt,krause1}
(cf. also the appendix of Ref. \cite{GS-jmp1}) the poloidal and
toroidal components of the $l$th angular and $n$th radial modes of
the multipole expanded fields $\bB_p^{(l,n)}$ and $\bB_t^{(l,n)}$
are given by $ \bB_p^{(l,n)}=
 -\nabla\times \left(\bbr\times\nabla \right)F_1^{(l,n)}(r,\theta),
 \ \bB_t^{(l,n)}=-\bbr\times\nabla F_2^{(l,n)}(r,\theta) $
with scalar functions $
F_{1,2}^{(l,n)}(r,\theta)=r^{-1}\Phi_{1,2}^{(l,n)}(r)Y_l^0(\theta)$
built from spherical harmonics $
Y_l^0(\theta)=\sqrt{(2l+1)/(4\pi)}P_l(\cos \theta) $ and the
solution components $\Phi_{1,2}^{(l,n)}$ of problem \rf{i7}. In our
case of only one BS solution the $n$-dependence reduces to a single
term $\Phi_{1,2}^{(l,1)}$. Expressing the components $F_\pm$ with
the help of Eqs. \rf{Af} and \rf{U} in terms of $\Phi_{1,2}^{(l,1)}$
\be{ma-22}
F_\pm=\pm\frac{r}{2\e}\left[\Phi_2^{(l,1)}-\left(\frac\alpha 2 \mp
\e\right)\Phi_1^{(l,1)}\right]
\ee
and using the inverted relations \cite{Moffatt} between the original
magnetic field strength $\bB^{(l,n)}$ and the corresponding scalar
functions $F_{1,2}^{(l,n)}$, $ \bbr\cdot\bB^{(l,n)}=
-l(l+1)F_1^{(l,n)}$, \ $\bbr\cdot(\nabla\times \bB^{(l,n)})=
-l(l+1)F_2^{(l,n)},\ $ we arrive for $F_-\equiv 0$ after multiplying
\rf{ma-22} with $-l(l+1)Y_l^0$ at
\be{ma-23}
-l(l+1)Y_l^0F_- =-\frac{r}{2\e}\bbr\cdot \left[\mu
\bj^{(l,1)}-\left(\frac\alpha 2 + \e\right)\bB^{(l,1)}\right]\equiv
0\,.
\ee
(The relation $\bB=\mu \bH$ and one of the Maxwell equations,
$\nabla\times \bH=\bj$, have been used.) Eq. \rf{ma-23} has to be
interpreted as special link between the induced current $\bj$ and
the spectrally shifted electromotive force $\alpha \bB$ in the
present dynamo model. However, it can be shown that the field is not
a Beltrami field $\nabla\times \bB\neq\beta \bB$.

We thank K.-H. R\"adler for useful comments on Beltrami fields.  BFS
is partially supported by grants RFBR-06-02-16719, SS-5103.2006.2.
He also thanks the Research Center Dresden-Rossendorf for
hospitality during his stays in Dresden and the financial support
under contract MHD 3-06. UG has been supported by the German
Research Foundation DFG, grant GE 682/12-3.

\end{document}